\documentclass[twocolumn]{aastex63}

\usepackage{graphicx}	
\usepackage{amsmath}	
\usepackage{amssymb}	

\submitjournal{ApJ}
\shorttitle{A QSO at z=6.62}
\shortauthors{Lu et al.}

\graphicspath{{./}{figures/}}

\begin{document}

\title{Subaru medium-resolution spectra of a QSO at z=6.62: Three reionization tests}

\correspondingauthor{Ting-Yi Lu}
\email{tingyilu@gapp.nthu.edu.tw}

\author[0000-0002-0786-7307]{Ting-Yi Lu}
\affiliation{Institute of Astronomy, National Tsing Hua University, 101, Sec. 2, Kuang Fu Rd., Hsinchu 30013, Taiwan}
\author[0000-0002-6821-8669]{Tomotsugu Goto}
\affiliation{Institute of Astronomy, National Tsing Hua University, 101, Sec. 2, Kuang Fu Rd., Hsinchu 30013, Taiwan}
\author[0000-0002-1860-0886]{Ji-Jia Tang}
\affiliation{Research School of Astronomy and Astrophysics, Australian National University, Cotter Road, Weston Creek, ACT 2611, Australia}
\affiliation{Graduate Institute of Astrophysics, National Taiwan University, No.1 Sec.4 Roosevelt Rd., Taipei 10617, Taiwan}
\affiliation{Institute of Astronomy and Astrophysics, Academia Sinica, No.1, Sec. 4, Roosevelt Rd., Taipei 10617, Taiwan}
\author[0000-0001-7228-1428]{Tetsuya Hashimoto}
\affiliation{Institute of Astronomy, National Tsing Hua University, 101, Sec. 2, Kuang Fu Rd., Hsinchu 30013, Taiwan}
\affiliation{Centre for Informatics and Computation in Astronomy, National Tsing Hua University, 101, Section 2. Kuang-Fu Road, Hsinchu, 30013, Taiwan}
\author{Yi-Hang Valerie Wong}
\affiliation{Institute of Astronomy, National Tsing Hua University, 101, Sec. 2, Kuang Fu Rd., Hsinchu 30013, Taiwan}
\author[0000-0002-9630-4003]{Chia-Ying Chang}
\affiliation{Institute of Astronomy, National Tsing Hua University, 101, Sec. 2, Kuang Fu Rd., Hsinchu 30013, Taiwan}
\author{Yi-Han Wu}
\affiliation{Institute of Astronomy, National Tsing Hua University, 101, Sec. 2, Kuang Fu Rd., Hsinchu 30013, Taiwan}
\author[0000-0001-9970-8145]{Seong Jin Kim}
\affiliation{Institute of Astronomy, National Tsing Hua University, 101, Sec. 2, Kuang Fu Rd., Hsinchu 30013, Taiwan}
\author[0000-0002-8560-3497]{Chien-Chang Ho}
\affiliation{Institute of Astronomy, National Tsing Hua University, 101, Sec. 2, Kuang Fu Rd., Hsinchu 30013, Taiwan}
\author{Ting-Wen Wang}
\affiliation{Institute of Astronomy, National Tsing Hua University, 101, Sec. 2, Kuang Fu Rd., Hsinchu 30013, Taiwan}
\author[0000-0003-4479-4415]{Alvina Y. L. On}
\affiliation{Institute of Astronomy, National Tsing Hua University, 101, Sec. 2, Kuang Fu Rd., Hsinchu 30013, Taiwan}
\affiliation{Centre for Informatics and Computation in Astronomy, National Tsing Hua University, 101, Section 2. Kuang-Fu Road, Hsinchu, 30013, Taiwan}
\affiliation{Mullard Space Science Laboratory, University College London, Holmbury St Mary, Surrey RH5 6NT, UK}
\author{Daryl Joe D. Santos}
\affiliation{Institute of Astronomy, National Tsing Hua University, 101, Sec. 2, Kuang Fu Rd., Hsinchu 30013, Taiwan}

\begin{abstract}

Investigating the Gunn-Peterson trough of high redshift quasars (QSOs) is a powerful way to reveal the cosmic reionization. As one of such attempts, we perform a series of analyses to examine the absorption lines observed with one of the highest redshift QSOs, PSO J006.1240+39.2219, which we previously discovered at $\mathnormal{z}$ = 6.62. Using the Subaru telescope, we obtained medium-resolution spectrum with a total exposure time of 7.5 hours.
We calculate the Ly$\alpha$ transmission in different redshift bins to determine the near zone radius and the optical depth at 5.6$<$\,$\mathnormal{z}$\,$<$6.5. We find a sudden change in the Ly$\alpha$ transmission at 5.75$<$\,$\mathnormal{z}$\,$<$5.86, which is consistent with the result from the literature. 
The near zone radius of the QSO is 5.79$\pm$0.09 $p$Mpc, within the scatter of the near  zone radii of other QSOs measured in previous studies.
We also analyze the dark gap distribution to probe the neutral hydrogen fractions beyond the saturation limit of the Gunn-Peterson trough. We extend the measurement of the dark gaps to 5.7$<$\,$\mathnormal{z}$\,$<$6.3. We find that the gap widths increase with increasing redshifts, suggesting more neutral Universe at higher redshifts. However, these measurements strongly depend on the continuum modeling. As a continuum model-free attempt, we also perform the dark-pixel counting analysis, to find the upper limit of $\langle$\,$x_{\rm H I}$\,$\rangle$ \,$\sim$0.6 (0.8) at $\mathnormal{z}<$5.8 ($\mathnormal{z}>$5.8). All three analyses based on this QSO show increasingly neutral hydrogen towards higher redshifts, adding precious measurements up to $\mathnormal{z}$\,$\sim$6.5.

\end{abstract}



\section{Introduction} \label{sec:intro}

Cosmic reionization is a major phase change of the Universe when the neutral intergalactic medium (IGM) was being ionized by the UV radiation from the first luminous objects \citep{Barkana2001,Loeb2001,Fan2006a}. It is crucial to uncover the reionization history and the early evolution of the ionizing sources. To date, the mid-point of the reionization has been constrained to be at $\mathnormal{z}$=7.7$\pm$0.7 from the CMB observation by \citet{Planck2018}.  High redshift quasars (QSOs) have been used as probes for the end of the reionization (see \citet{Mortlock2016} for a review).
By measuring the Ly$\alpha$ transmissions at different redshifts in the UV continuum of a QSO and therefore the Gunn-Peterson (GP) optical depths, the end of the reionization at that line-of-sight (LOS) can be revealed.  Previously, the occurrences of the GP troughs (where the Ly$\alpha$ absorptions in the spectra of QSOs are saturated) through different LOSs have been reported to be at $\mathnormal{z}$\,$\sim$6  \citep[e.g.][]{Becker2001,Songaila2004,Fan2006b, Goto2011,Bolton2011,Becker2015,Eilers2017,Eilers2018,Bosman2018}, indicating the end stage of the reionization. On the other hand, because the GP optical depth is very sensitive to the fractional change of the scarcely remaining neutral hydrogen (neutral fraction, $x_{\rm H I}$\,$\sim$\,$10^{-4}$) during that stage \citep{Gunn1965, Fan2006a}, the occurrence of the GP trough limits our capability in probing the end-stage of the reionization toward $\mathnormal{z}$\,$\gtrsim$6, where the average $x_{\rm H I}$ becomes higher than $10^{-4}$. In addition, current simulations \citep[]{Nasir2019,Keating2019,Keating2020} suggest that the scatter of GP optical depth  \citep[]{Fan2006b, Becker2015,Eilers2018,Bosman2018} could be explained by late reionization model as well. Thus, to advance the research, other analyses on the $\mathnormal{z}$\,$\gtrsim$6 QSOs are required.
\\
One of the alternative, higher order approaches to measure reionization at high redshift is to measure the frequency of high Ly$\alpha$ transmission peaks in the spectra ~\citep{Croft1998, Songaila2002}. Such transmission peaks can trace the higher ionization state regions in the GP trough. The practical measurement on the frequency is performed by counting the dark gaps, which is the gap constructed by two adjacent peaks.   \citet{Songaila2002} first illustrated the distribution of the dark gap in QSO spectra at 3.5$<\mathnormal{z}<$5.5 using a sample of 15 QSOs at 4.42$\le$\,$\mathnormal{z}$\,$\le$5.75; ~\citet{Paschos2005} generated the synthetic spectra of the Ly$\alpha$ forest up to $\mathnormal{z}$\,$\sim$6.6 and performed the simulation of the dark gap distribution in the spectra at $\mathnormal{z}$=5.5-6.0 and $\mathnormal{z}$=6.0-6.5; ~\citet{Gallerani2006} performed the dark gap simulation at similar redshift with larger box size; and ~\citet{Fan2006b} extended the dark gap measurement to $\mathnormal{z}$\,$\sim$6 with a sample of 19 QSOs. These studies suggested that the gap width may have a notable change at $\mathnormal{z}$\,$\sim$6. More recently, ~\cite{Gnedin2017} present the comparison between the simulations from the Cosmic Reionization On Computers project and the observations of the QSOs at 5.7$<$\,$\mathnormal{z}$\,$<$6.4 from ~\cite{Fan2006b} and ~\citet{Becker2015}. The shape of the transmission peaks may also be a sensitive probe of the intergalactic medium (IGM) environment.  
In addition to the above two methods which are dependent on the assumption of intrinsic UV continuum, \citet{McGreer2011} provided a model-free constraint on the upper limit of the volume-averaged neutral fraction by counting the covering fraction of dark pixels, the pixels with no flux detected. However, these analyses are more demanding on the signal-to-noise ratio (S/N) of the data to minimize the impact of the noise on the low transmission region.
\\
So far, $\sim$50 QSOs at $\mathnormal{z}$\,$>$6.5 are discovered (see \citet{Inayoshi2019} for a full list of the 203 known $\mathnormal{z}$\,$>$6 QSOs). While only $\lesssim$10 of them are used to constrain the reionization \citep{Tang2017,Bosman2018,Eilers2018,Banados2018} due to the expensiveness of obtaining high quality spectra. To push the constraint to $\mathnormal{z}$\,$\gtrsim$6.5  statistically, more sight-lines  through the corresponding redshift should be observed. 
\\
In this paper, we carry out a follow-up investigation on PSO J006.1240+39.2219 ($\mathnormal{z}$=6.621) with the Subaru Faint Object Camera and Spectrograph  \citep[FOCAS; ][]{Kashikawa2002}. 
We observe the medium-resolution UV spectrum of this QSO in order to add a sight-line to the $\mathnormal{z}$\,$\gtrsim$6.5 sample to constrain reionization at high redshift. In addition, since this QSO is a super-Eddington QSO \citep{Tang2019}, it is of interest to study its physical properties via emission lines. We present the line luminosities in Section~\ref{sec:Analyses}.  
The IGM transmission, the GP optical depth and the near zone radius of this QSO which will help constraining the IGM neutral fraction are presented in Section~\ref{sec:GPtest}  
Furthermore, as we take advantage of the high S/N optical spectrum, we are able to perform dark gap statistics to investigate the subtle change of the structure in the spectrum  (Section~\ref{sec:dg}), and the dark pixel test, which is a nearly model-independent measurement of the volume-averaged neutral fraction (Section~\ref{sec:dp}) for the first time to $\mathnormal{z}$\,$\sim$6.5. 
\\
Throughout the paper, we use a $\Lambda$CDM cosmology with $\text{H}_{0}$=70 km $\text{s}^{-1} \text{Mpc}^{-1}$, $\Omega_{m}$=0.3 and $\Omega_{\Lambda}$=0.7.

\section{Data}
\label{sec:data}
\begin{figure*}
	\includegraphics[width=\textwidth]{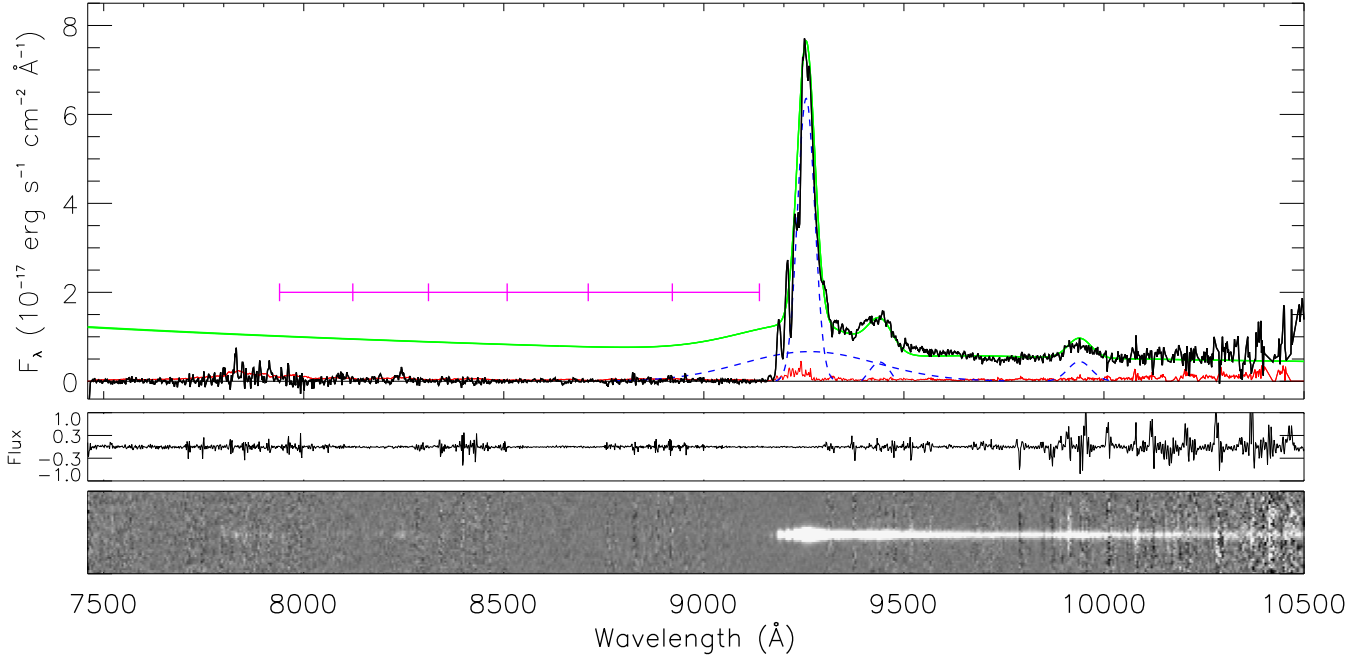}
    \caption{The Subaru/FOCAS spectrum of PSO J006.1240+39.2219. In the upper panel, the green line is the best-fit continuum. The blue dashed lines are Gaussian fits to the emission lines. The redshift bins shown in purple horizontal line are used in analyzing the transmissions and optical depths. The red lines show the measured errors. The middle panel shows the sky spectrum in an arbitrary scale. The bottom panel shows the 2D spectrum of the QSO.} 
    \label{spec}
\end{figure*}

In previous work we discovered PSO J006.1240+39.2219 ~\citep{Tang2017}, yet the spectrum was of low resolution and a short exposure time of 5000 seconds, the S/N at 1250\AA$<\lambda_\text{rest}<$1280\AA \, was only 9.82. Therefore, we carry out additional spectroscopic observation of PSO J006.1240+39.2219 with the Subaru/FOCAS using the medium-resolution VPH950 grism with 058 blocking filter (PI: Goto). The exposure time is 27,000 seconds, or 7.5 hours. The wavelength coverage is from 7500 to 10450 \AA. ~The spectral resolution with the 0.4'' slit is $R\sim$5500. 
The information of this QSO is summarized in Table~\ref{qso_info}. We adopt the redshift calculated by ~\citet{Mazzucchelli2017} based on  [C~{\sc ii}]\,158$\mu$m line. The spectrum of the QSO we obtained is shown in Fig.~\ref{spec}. 

Data reduction is performed in a standard manner using {\em IRAF}. After flat-fielding, the wavelength calibration is performed using skylines. We perform first-order sky background subtraction using 60 pixels on both sides of the QSO position in the spatial direction. We chose the sky regions as close to the QSO as possible, but at the same time avoiding the tail of the QSO flux.
 After subtracting the background, the 1-$\sigma$ clipped median flux bluewards of the QSO's near zone region is 1.15$\pm$6.33$\times 10^{-19}$erg cm$^{-2}$s$^{-1}$\AA$^{-1}$, which is consistent with zero. The median S/N at 1250\AA$<\lambda_\text{rest}<$1280\AA\, is 17.95.
We observe BD+28D4211 as a standard star for 90 seconds. We apply the same trace as our standard star because the QSO flux bluewards of Ly$\alpha$ is significantly absorbed. This star is used for flux calibration. The 1D extracted spectrum is then binned into $\sim$1.5 \AA\, pixel$^{-1}$. 

\begin{table*}
	\centering
	\caption{Properties of PSO J006.1240+39.2219.}
	\label{qso_info}
	\begin{tabular}{lcccccr} 
		\hline
		Object &R.A. (J2000)&DEC. (J2000)& $\mathnormal{z}$\,$_{[\text{\ion{C}{2}}]}$  & R$_\text{p,NZ}$ (pMpc) & R$_\text{NZ,cor}$ (pMpc) \\
		\hline
		PSO J006.1240+39.2219 & 00:24:29.657&+39:12:00.66&6.621$\pm$0.002$^{a}$ &4.19$\pm$0.06 &5.79$\pm$0.09\\
		\hline
	\end{tabular}
  \\
  $^{a}$ Reference: ~\citet{Mazzucchelli2017}
\end{table*}

\section{Analyses}
\label{sec:Analyses}

For the analysis in the later sections, we use the power-law equation $F_{\lambda}$=$F_\text{0}$\,$(\lambda/2500\text{\AA})^{\alpha_{\lambda}}$ \citep{Mazzucchelli2017} to fit the continuum. We adopt the slope $\alpha_{\lambda}$ = -2.94$\pm$0.03 fitted to the rest-frame wavelength windows [1285-1295; 1315-1325; 1340-1375; 1425-1470; 1680-1710; 1975-2050; 2150-2250; and 2950-2990] \AA\,  of the near-infrared spectrum from \citet{Tang2019}. The $F_{0}$ fitted to the wavelength window is 5.40 $\pm$0.05 10$^{19}$ erg s$^{\text{--}1}$\, $\text{cm}^{-2}  \text{\AA}^{-1}$. 
The windows are selected to avoid the strong emission lines.\\
We also estimate the luminosity of the Ly$\alpha$, $N_{\rm V}$,$\lambda$1239,1243 and $O_{\rm I}$\,$\lambda$1304+\,$Si_{\rm II}$\,$\lambda$1306 emission lines by fitting double Gaussian plots. The results are shown in Fig.~\ref{spec} and Table~\ref{lum}.

\begin{table}
	\centering
	\caption{The Ly$\alpha$, \ion{N}{5}\,$\lambda$1239,1243 and \ion{O}{1}\,$\lambda$1304+\,\ion{Si}{2}\,$\lambda$1306 emission line luminosities.}
	\label{lum}
	\begin{tabular}{lr} 
		\hline
		Line & Luminosity (10$^{44}$ erg s$^{\text{--}1}$) \\
		\hline 
        Ly$\alpha$ & 17.75$\pm$0.10\\
		\ion{N}{5}\,$\lambda$1239,1243& 1.79$\pm$0.06\\
		\ion{O}{1}\,$\lambda$1304+\,\ion{Si}{2}\,$\lambda$1306 & 1.54$\pm$0.16\\
		\hline
	\end{tabular}
\end{table}

\subsection{IGM optical depth}
\label{sec:GPtest} 
\begin{figure}
	\includegraphics[width=\columnwidth]{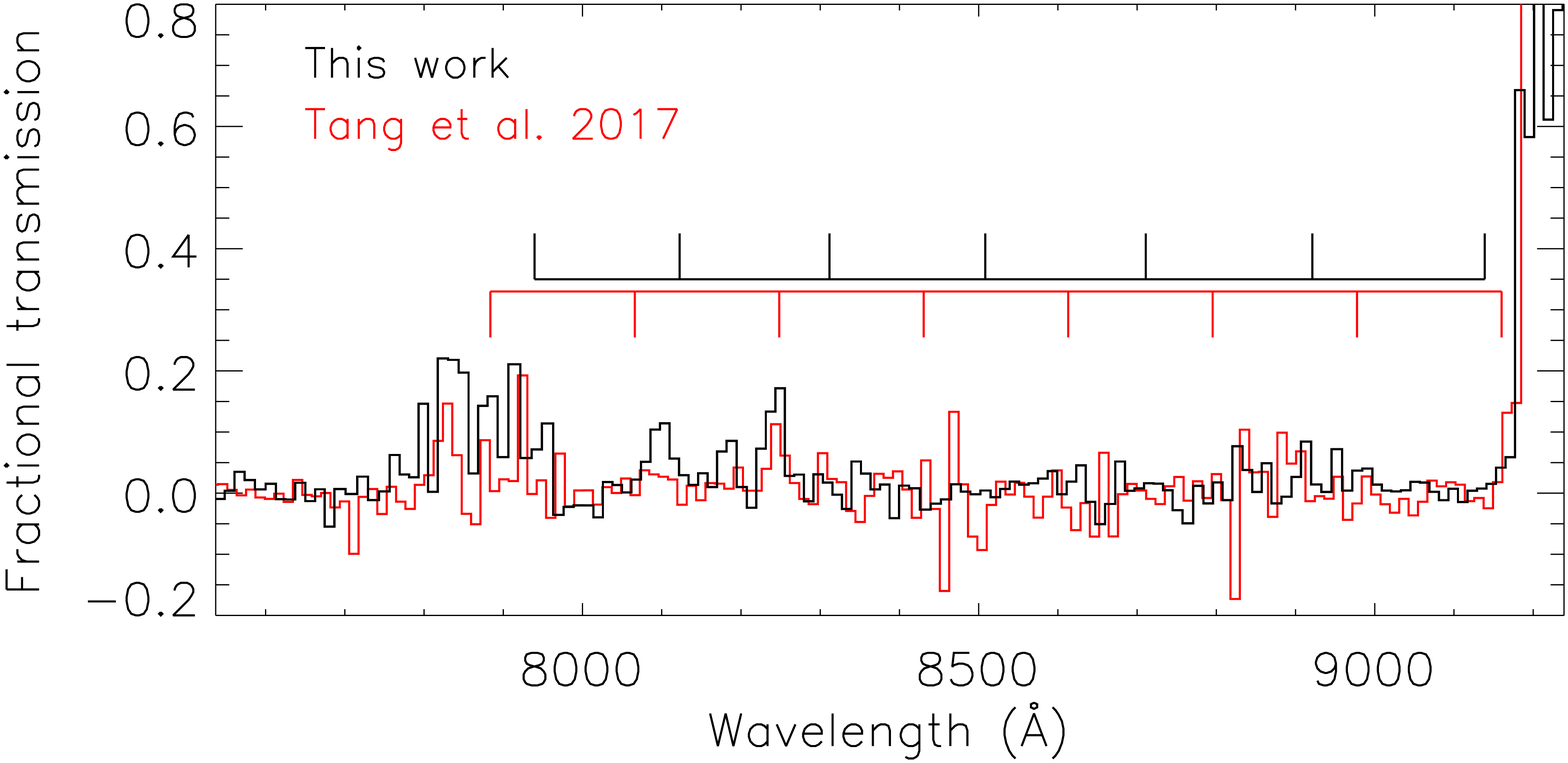}
    \caption{Fractional transmission spectra of this work (black) and the spectrum taken from \protect\citet{Tang2017} (red). Both are 8-pixel binned. The horizontal black and red rulers show the redshift bins used in this work and in \protect\citet{Tang2017}, respectively. }
    \label{fractrans}
\end{figure}

\begin{figure}
    \includegraphics[width=\columnwidth]{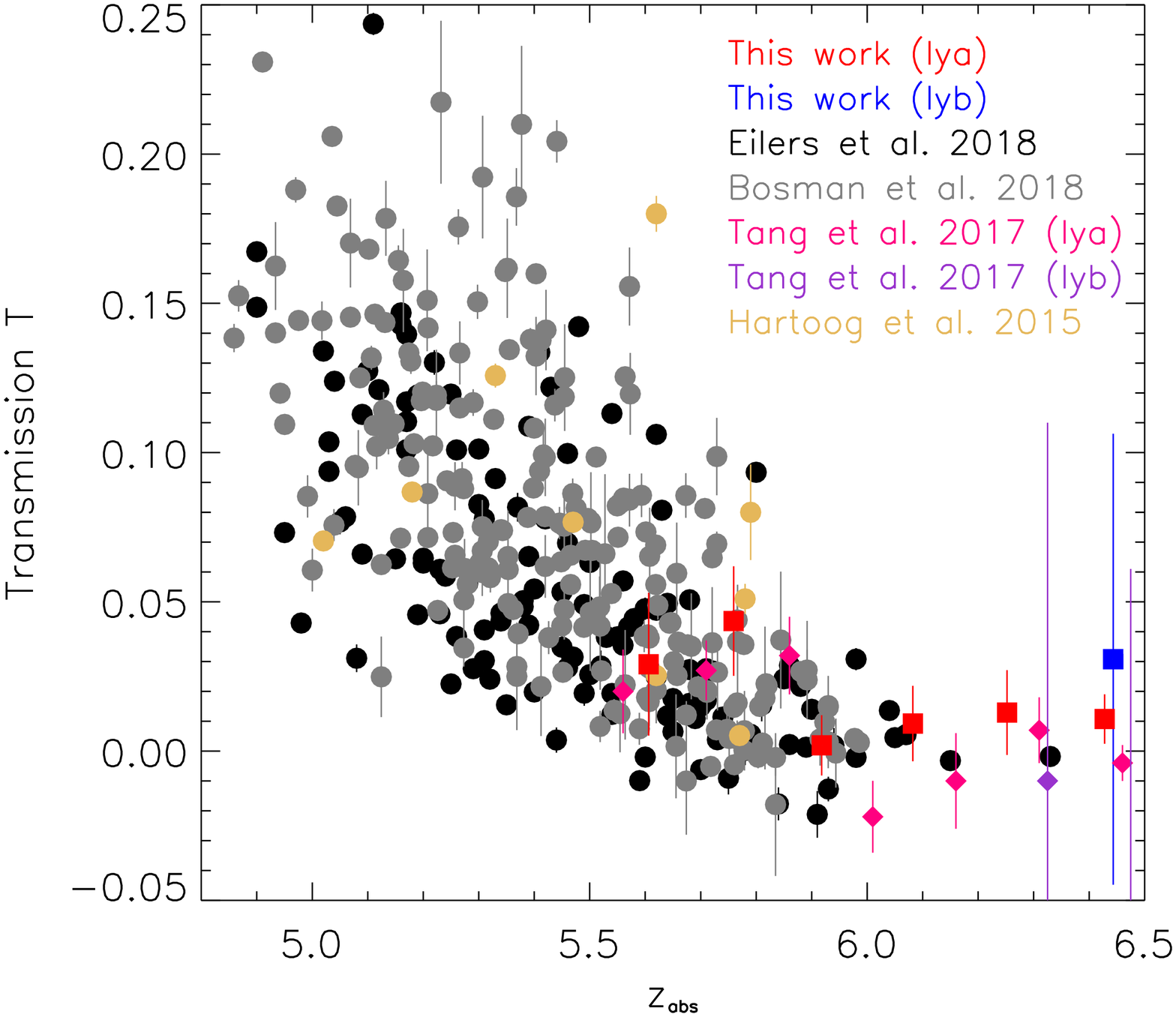}
    
    \caption{Ly$\alpha$ transmission versus redshift. Red and blue squares with error bar are the Ly$\alpha$ and the effective Ly$\beta$  transmission in the spectrum of PSO J006.1240+39.2219, respectively. Pink and purple diamonds with error bar are the result of the same QSO from \protect\citet{Tang2017}. The rest of the data points are from previous studies \protect\citep{Hartoog2015, Bosman2018,Eilers2018} as indicated in the legend. 
    Data points from \protect\citet{Tang2017} are measured through the same LOS as this work, while \protect\citet{Hartoog2015} is measured from a gamma-ray burst spectrum.
    }
    \label{trans}
\end{figure}

\begin{figure}
    \includegraphics[width=\columnwidth]{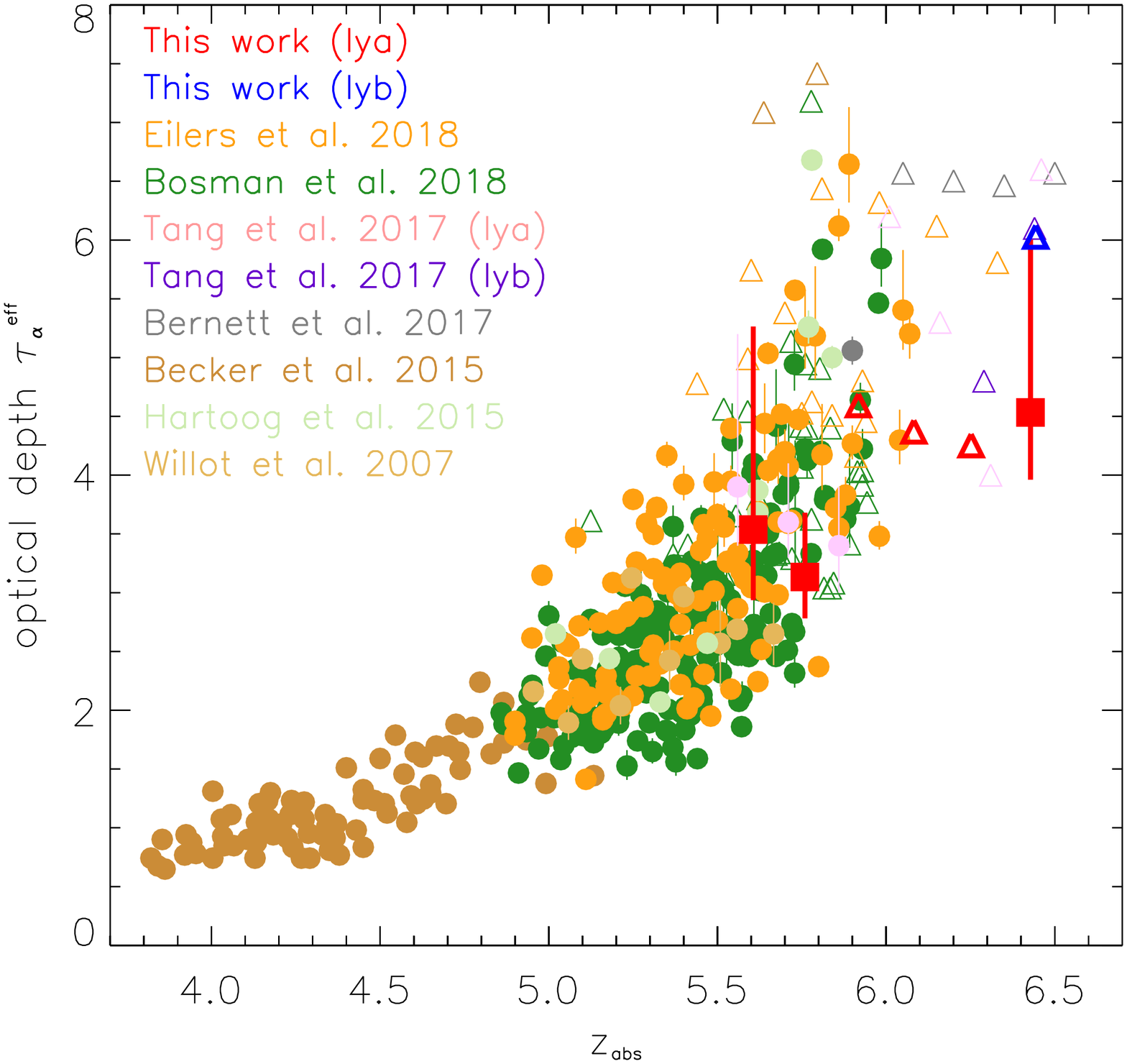}
    \caption{Effective Ly$\alpha$ optical depth as a function of redshift. Red (Ly$\alpha$) and blue (Ly$\beta$) squares with error bar or triangles are results from this work. The other data are taken from previous studies \protect\citep{Willott2007, Becker2015, Hartoog2015, Barnett2017, Tang2017, Bosman2018, Eilers2018}. The triangles are the lower limit of the optical depth.}
    \label{tau}

\end{figure}

To investigate neutral hydrogen absorption as a function of redshift, we follow various methods from previous studies \citep[e.g.][]{Fan2002, Songaila2004,  Becker2015, Tang2017, Bosman2018} to perform a Gunn-Peterson test on the QSO with the fitted power-law continuum plus Ly$\alpha$ emission as the intrinsic flux. In this test, the steep rise of the average transmission can be considered as a sign of the end of the reionization.
\\
We measure the average transmission of the Ly$\alpha$, the weighted Ly$\beta$ transmissions, and the optical depths at 7939 - 9139\AA \,(Ly$\alpha$) and 7548-7732\AA \,(Ly$\beta$). 
We select these wavelength intervals in order to avoid the Ly$\alpha$ emission line, the near zone region, and the Ly$\beta$ emission line. But the highest redshift bin could be contaminated by the near zone. The wavelength range is divided into 50 cMpc/h bins. Aside from that, we masked the wavelengths with strong skyline residuals while measuring transmission.
\\
The Ly$\alpha$ transmission (\textit{T}) is measured as \textit{T} = $\langle$\,$F_{\nu\text{,obs}}$/$F_{\nu\text{,int}}$\,$\rangle$ , and the optical depths ($\tau$) are derived from the transmissions, where $\tau$=$-\ln{(T)}$ \citep{Fan2006b}. $F_{\nu\text{,obs}}$ is the observed flux and  $F_{\nu\text{,int}}$ is the intrinsic continuum. Our transmission spectrum and the transmission spectrum from \citet{Tang2017} are shown in Fig.~\ref{fractrans} for comparison. 
Since the Ly$\beta$ forest overlaps with the Ly$\alpha$ forest in the spectrum, and the Ly$\beta$ optical depths are a factor of several smaller than the Ly$\alpha$ optical depths for the same hydrogen density, we calculate the weighted Ly$\beta$ transmissions and their effective optical depths ($\tau_{\beta}^\text{eff}$). Following the discussion in \citet{Fan2006b}, we first calculate the weighted Ly$\beta$ transmissions considering the foreground Ly$\alpha$ transmission at the corresponding observed wavelength. The foreground Ly$\alpha$ transmission is calculated using Equation (5) in \citet{Fan2006b}, which is the best-fit power law for the optical depth at $\mathnormal{z}$\,$<$5.5. Then we convert the Ly$\beta$ transmissions into the effective optical depths by applying $\tau_{\alpha}$/$\tau_{\beta}$=2.25 from \citet{Fan2006b}. 
The result is shown in Table~\ref{trans_tau}, and is plotted in Fig.~\ref{trans} and Fig.~\ref{tau}.
\\
The uncertainty of the continuum and the noise level are taken into account in estimating the transmission error. When the flux detection is less than 2-$\sigma$, we estimate the lower limit of the corresponding optical depth using 2-$\sigma$ error of the transmission.
\\

\begin{table}
	\centering
	\caption{The transmissions and optical depths of PSO J006.1240+39.2219. The transmission error and the optical depth error are in 2-$\sigma$ error. }
	\label{trans_tau}
	\begin{tabular}{lcccr} 
		\hline
		Redshift &Wavelength range& Line & Transmission & $\tau_{\alpha}$\\
		\hline 
		6.43 &8921-9139& Ly$\alpha$ & 0.011$\pm$0.008 & 4.6$^{+1.5}_{-0.6}$\\
		6.25 &8711-8921& Ly$\alpha$ & 0.013$\pm$0.014 & $>$4.6\\
		6.08 &8508-8711& Ly$\alpha$ & 0.009$\pm$0.013 & $>$4.4\\
		5.92 &8312-8508& Ly$\alpha$ & 0.002$\pm$0.010 & $>$4.6\\
		5.76 &8123-8312& Ly$\alpha$ & 0.043$\pm$0.018 &  3.1$^{+0.5}_{-0.4}$\\
		5.61 &7939-8123& Ly$\alpha$ & 0.029$\pm$0.024 &  3.5$^{+1.7}_{-0.6}$\\
		\hline
		6.45 &7548-7732& Ly$\beta$ & 0.031$\pm$0.075 & $>$6.0\\
		\hline
	\end{tabular}
\end{table}

\begin{figure}
    \includegraphics[width=\columnwidth]{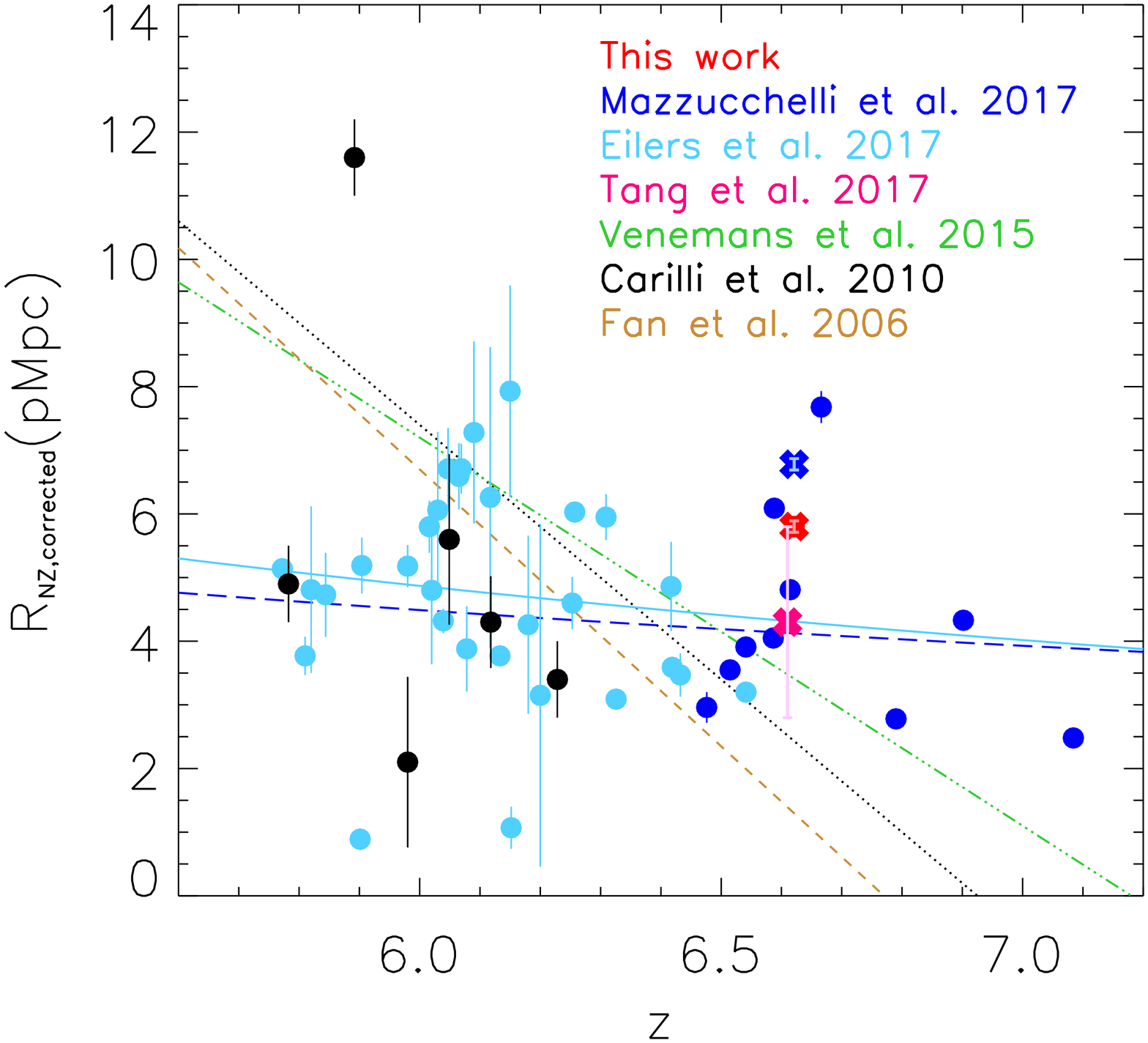}
    \caption{$R_\text{NZ,corrected}$ of the QSOs. The red cross with error bar is the result from this study. The brown dashed line is the fitted trend of $R_\text{NZ,corrected}$ from \protect\citet{Fan2006b}; the green dash dot line is from \protect\citet{Venemans2015}; the black dots and dotted line is from \protect\citet{Carilli2010}; the cyan dots and solid line are from \protect\citet{Eilers2017}; the blue dots and long dash line are from \protect\citet{Mazzucchelli2017}. The pink and blue crosses with error bar are the $R_\text{NZ,corrected}$ of the same QSO as which of this work measured by \protect\citet{Tang2017} and \protect\citet{Mazzucchelli2017}.}
    \label{fig:rpnz}
\end{figure}

We also measure the near zone radius ($R_\text{p,NZ}$) of the QSO, using $R_\text{p,NZ}$= ($D_\text{Q}$-$D_\text{GP}$)/(1+$\mathnormal{z}_\text{Q}$) \citep{Fan2006b}. We also correct the near zone radius for the luminosity difference, $R_\text{NZ,corrected}$= $R_\text{p,NZ}$\,$\times$\,$10^{0.4(27+M_{1450})/3}$ \citep{Carilli2010}. $D_\text{Q}$\, and $D_\text{GP}$ stand for the comoving distance of the QSO and where the transmission first drops below 0.1, respectively. The near zone measurement is performed after smoothing the spectrum to R$\sim$2600 with a boxcar function, following previous studies \citep[e.g.][]{Venemans2015, Eilers2017}. We measure the values of $R_\text{p,NZ}$ and $R_\text{NZ,corrected}$ to be 4.19$\pm$0.06 and 5.79$\pm$0.09 ($p$Mpc), respectively. The uncertainty of the near zone radius is $\Delta$\,$R_\text{p,NZ}$ $\sim$0.1 ($p$Mpc), due to the $\mathnormal{z}_\text{[\ion{C}{2}]}$ uncertainty of $\Delta$\,$\mathnormal{z}$\,$\sim$0.002 \citep{Carilli2010}.  These values are also shown in Table~\ref{qso_info} and Fig.~\ref{fig:rpnz} with comparison to previous literature results.\

\subsection{Dark gap statistics}
\label{sec:dg}
\begin{figure*}
	\includegraphics[width=17cm]{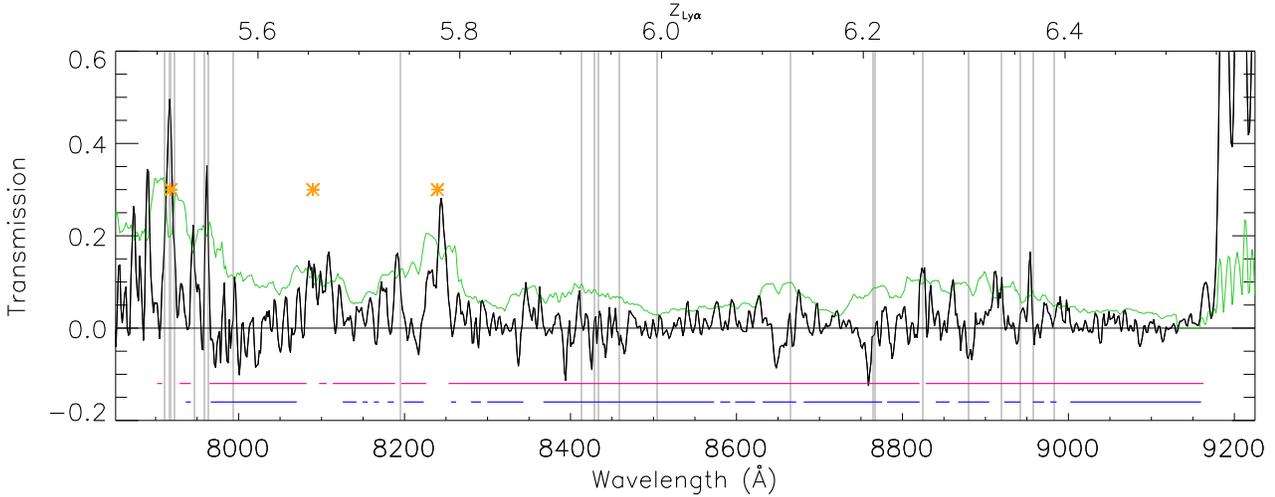}
    \caption{The GP trough in the QSO spectrum. The spectrum was smoothed to R$\sim$2600. The green lines are the 2-$\sigma$ spectral error. The orange asterisks show the positions of transmission spikes with more than 4 pixels with flux level higher than the spectral error. The upper pink horizontal lines mark the \protect$\tau>$2.5 dark gaps. The lower blue horizontal lines mark the \protect$\tau>$3.5 dark gaps. The gray vertical lines show the positions of the pixels with strong skyline residuals. }
    \label{fig:dgspec}
\end{figure*}

\begin{figure}
    \includegraphics[width=\columnwidth]{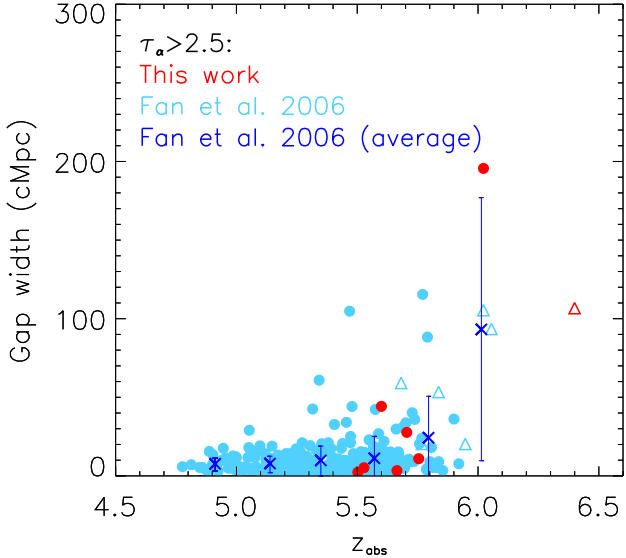}
    \caption{The $\tau>2.5$ dark gap sizes and their corresponding redshift. The red dots and triangle are from this work. The light blue dots and triangles are the gap sizes taken from \protect\citet{Fan2006b}. Both red and light blue triangles represent the lower limit of the gap sizes. The deep blue crosses with error bar are the average gap sizes of the \protect\citet{Fan2006b} data.}
    \label{fig:gap25}
\end{figure}

\begin{figure}
    \includegraphics[width=\columnwidth]{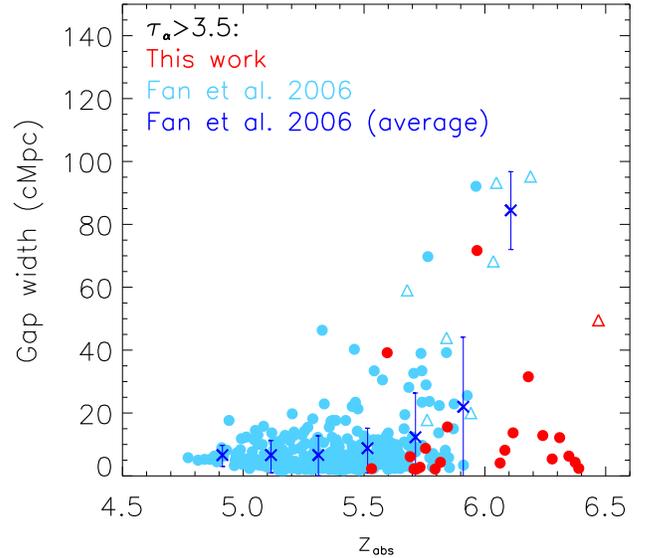}
    \caption{The $\tau>3.5$ dark gap sizes and their corresponding redshift. All the legends are the same as Fig.~\ref{fig:gap25}.}
    \label{fig:gap35}
\end{figure}

\begin{figure*}
	\includegraphics[width=\textwidth]{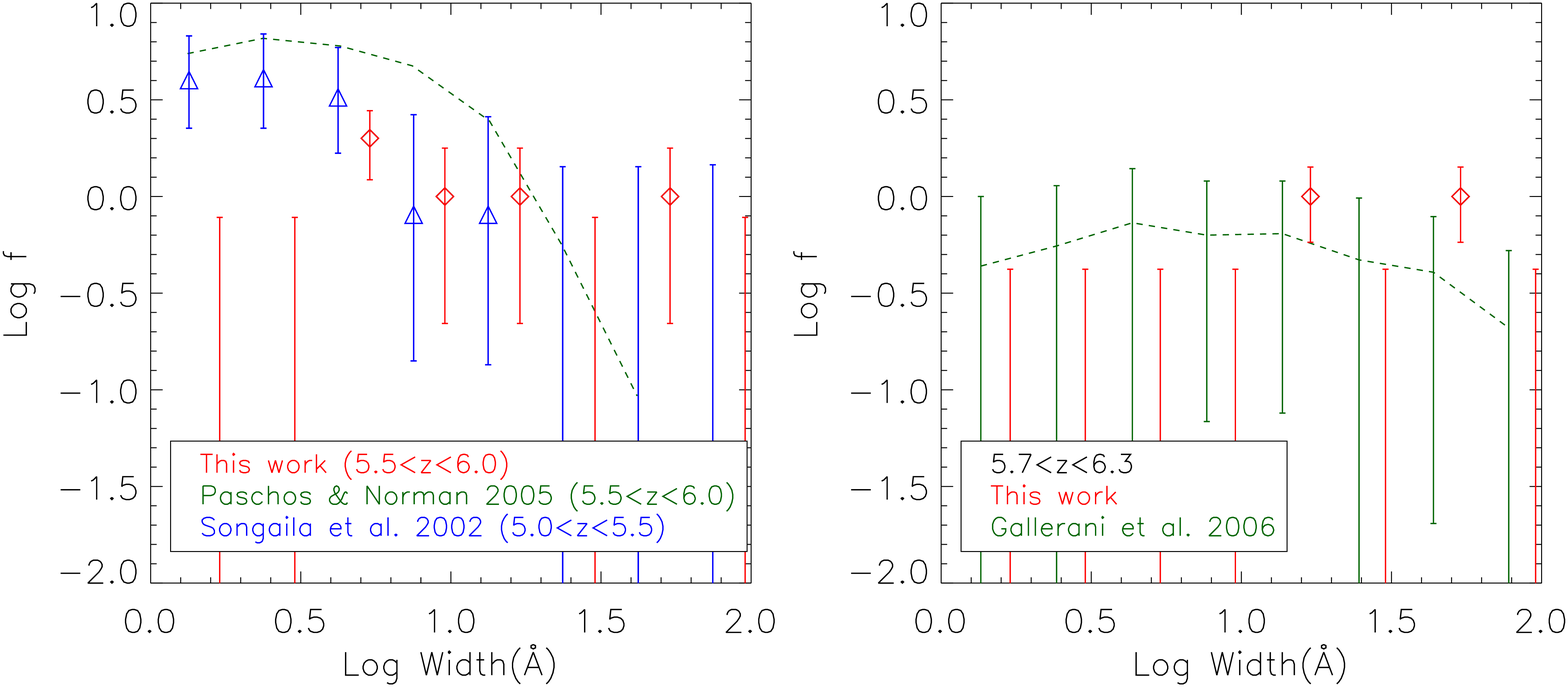}
    \caption{The number of dark gaps ($\tau>$2.5) per unit redshift versus the gap width. Left: Red diamonds are the distribution of gaps from this work (5.5$<$z$<$6.0). Blue triangles are the observational result (5.0$<$z$<$5.5) taken from \protect\citet{Songaila2002}. The green dashed line are the simulation of \protect\citet{Paschos2005}. Right: Red diamonds are the distribution of gaps from this work (5.7$<$z$<$6.3). The green dashed line are the simulation of \protect\citet{Gallerani2006}.}
    \label{fig:cpsgpa}
\end{figure*}

 With increasing redshift, absorption increases and it becomes more difficult to detect flux. Hence, studying the epoch of reionization with the evolution of the average transmission may become less powerful. In contrast, measuring the separation between the sparse remaining flux becomes a better approach \citep{Songaila2002,Becker2005,Fan2006b,Gallerani2006,Gallerani2008,Gnedin2017}. With the medium-resolution spectrum, we can measure the dark gaps statistics for the first time at $\mathnormal{z}$\,$\sim$6.5 with PSO J006.1240+39.2219.
The wavelengths with detected flux in the spectrum could be seen as the relatively ionized regions where the light was not fully absorbed by the hydrogen. Namely, measuring the intervals of those peaks in the spectrum could be a probe of the occurrence frequency of the ionized regions. 
We follow the discussion in Section 5 of ~\citet{Songaila2002}, measuring the width of the gap which is constructed by two adjacent, $\tau$\,$\leq$2.5 peaks, and then count the number of the gaps within the corresponding gap width bins in a redshift interval after smoothing the spectrum to R$\sim$2600. We discard the peaks with the width narrower than the smoothed resolution, since the widths of the sky spikes at the observed frame are mostly narrower than the smoothed resolution. However, note that there could remain sub-splitting of the opaque regions by wider skyline residuals. 
We also measure the gaps between the $\tau$\,$\leq$3.5 peaks since the neutral fraction is higher at $\mathnormal{z}>6$, and the $\tau >$2.5 dark gap extends to the near zone and can only be the lower limit of the gap width in many cases \citep{Fan2006b}. The gap positions are shown in Fig.~\ref{fig:dgspec}. The development of the $\tau>$2.5 and the $\tau>$3.5 gaps are shown in Figs.~\ref{fig:gap25} and ~\ref{fig:gap35}, respectively, and the redshifts and widths of the gaps are shown in Table ~\ref{darkgaptab}. The distributions of the  $\tau>$2.5 gaps at 5.5$<\mathnormal{z}<$6.0, and 5.7$<\mathnormal{z}<$6.3 are shown in Fig.~\ref{fig:cpsgpa}.

\begin{table*}
	\centering
	\caption{The redshifts and widths of the $\tau_{\alpha}>2.5$ and $\tau_{\alpha}>3.5$ dark gaps in the spectrum of PSO J006.1240+39.2219. }
	\label{darkgaptab}
	\begin{tabular}{lccr} 
		\hline
		Optical depth&Redshift range&z$_{average}$&gap length (cMpc)\\
		\hline 
		$\tau_{\alpha}>2.5$&6.26-6.54&6.40&106.69\\
	                 	&5.79-6.26&6.02&195.60\\
                         &5.74-5.77&5.75&10.96\\
                         &5.67-5.74&5.71&27.70\\
                         &5.66- 5.67&5.66& 3.35\\
                         &5.55-5.65& 5.60&44.24\\
                         &5.52-5.53&5.53&5.19\\
                         &5.50-5.50& 5.50&2.32\\
		\hline
         $\tau_{\alpha}>3.5$&6.41-6.53&    6.47 &      49.52 \\
                         &6.39-6.39&   6.39   &    2.40   \\
                         &6.37- 6.38&  6.37     &  4.33   \\
                         &6.34- 6.36&     6.35     & 6.28\\
                         &6.29-6.32&6.31   &    12.18\\
                         &6.27-6.29&    6.28     &  5.39 \\
                         &6.22- 6.26& 6.24       &12.85\\
                         &6.14- 6.22& 6.18   &    31.52\\
                         &6.10- 6.13&   6.11&       13.69\\
                         &6.07-6.09&  6.08&       8.17  \\
                         &6.06-6.07&   6.06 &      4.10 \\
                         &5.88-6.05&   5.97&       71.69\\
                         &5.83-5.86&    5.85&      15.58 \\
                         &5.81-5.82&   5.82 &     4.33\\
                         &5.79-5.80&    5.79 &      2.17\\
                         & 5.74-5.76&   5.75   &    8.77\\
                         &5.73-5.73&   5.73   &   2.75\\
                         &5.71-5.72&   5.72   &   2.21\\
                         &5.70-5.71&   5.71    &   2.21\\
                         &5.68-5.70&5.69   &    6.11\\
                         &5.55-5.64& 5.60  &   39.17\\
                 		&5.53-5.53&5.53&2.30\\
		\hline
	\end{tabular}
\end{table*}


\subsection{Dark pixel}
\label{sec:dp}
\citet{McGreer2011} and \citet{McGreer2015} proposed a model-independent method to constrain the upper limit of the volume-average neutral hydrogen fraction ($\langle$\,$x_{\rm H I}$\,$\rangle$) by directly measuring the covering fraction of the dark pixels. The dark pixels are the pixels in the spectrum with flux under a given threshold, which was set to be 2-$\sigma$ root-mean square (rms) noise \citep{McGreer2011} or zero flux \citep{McGreer2011, McGreer2015}. Two cases resulting in the dark pixels are discussed in \citet{McGreer2015}: the pre-reionization neutral hydrogen in any physical region and the collapsed self-shielded region cause saturated absorption (the observed flux under the threshold); the ionized gas with sufficient optical depth result in fewer transmitted flux, which is below the detection limit. Compared to the neutral fraction derived from the IGM optical depth measurement, which requires assumptions on the intrinsic QSO continuum and uniform background, the dark pixels measurement provide a more robust result of the the neutral fraction.
\\
The only assumption made in this measurement is that the dark pixels are distributed symmetrically around the threshold. In addition, it requires high S/N ratio spectrum, in order to reduce the probability that a dark pixel counted as bright pixel (pixel which is not dark) due to the noise.

\begin{figure}
	\includegraphics[width=\columnwidth]{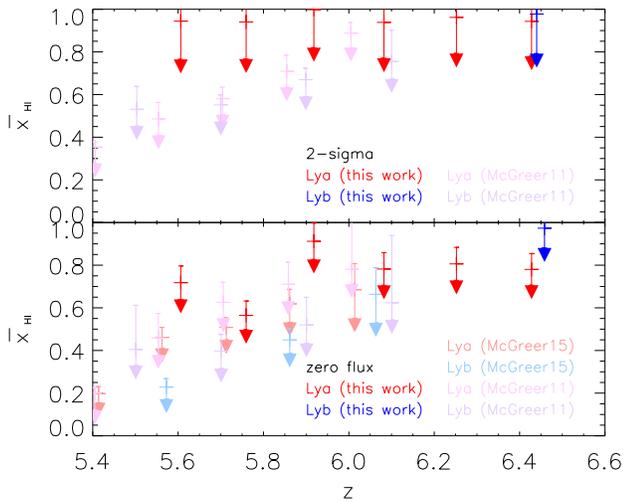}
    \caption{The upper limit of  $\langle$\,$x_{\rm H I}$\,$\rangle$. The upper panel shows the test using 2-$\sigma$ threshold. The red and blue arrows are the results from this work. The length of arrows represents the statistical error; The pink and purple arrows are taken from \protect\citet{McGreer2011}. The lower panel shows the test using the zero flux threshold. The red and blue arrows are from this work; The pinks and purples are from \protect\citet{McGreer2011}; The faint red and blue arrows are from the data of \protect\citet{McGreer2015}. }
    \label{fig:dps}
\end{figure}

We follow the procedure from \citet{McGreer2011}. First, we divided the sky-masked spectrum into bins of pixels with a width of 3.3 pMpc \citep[see the discussion in Section 3.3 of ][]{McGreer2011}. Then we count the covering fraction of dark pixels where the flux $\leq$2-$\sigma$ (2-$\sigma$ threshold), and where the flux $\leq$0 (zero flux threshold) in the same six 50 cMpc/h redshift bins as which in Sec. \ref{sec:GPtest}. 
For the 2-$\sigma$ threshold, $\langle$\,$x_{\rm H I}$\,$\rangle$ are then derived by scaling the dark pixel covering fractions by a factor of 1.023 because 2.3 per cent of the dark pixels are expected to scatter above the threshold, assuming Gaussian statistics. The uncertainties of the $\langle$\,$x_{\rm H I}$\,$\rangle$ are calculated after scaling. In contrast to the way the uncertainties are calculated in \cite{McGreer2011, McGreer2015}, where the jackknife method was adopted for the multi sight lines, we simply calculate the uncertainties by taking the square root of the fractions for single sight line. For the zero flux threshold, the scaling factor is adjusted to be 2, because the dark pixels under this threshold are expected to scatter equally about zero. The uncertainties for the zero flux threshold are calculated in  the same manner as for the 2-$\sigma$ threshold. Results are shown in Fig.~\ref{fig:dps} and Table~\ref{darkpixtab}.
The $\langle$\,$x_{\rm H I}$\,$\rangle$ measured using both criteria shows growing trend toward high redshift, but the result from the 2-$\sigma$ threshold shows higher  $\langle$\,$x_{\rm H I}$\,$\rangle$. This difference is expected since at such redshifts, the transmitted fluxes approach zero. Hence, using the zero flux threshold is more appropriate because it could reveal subtler fractional changes than using the 2-$\sigma$ threshold.

\begin{table*}
	\centering
	\caption{The dark pixel fractions in each redshift bins measured with 2-$\sigma$ criteria of with zero flux criteria. }
	\label{darkpixtab}
	\begin{tabular}{lcccr} 
		\hline
		Redshift &Wavelength range (\AA)& Line & $\langle$\,$x_{\rm H I}$\,$\rangle$ (2-$\sigma$) & $\langle$\,$x_{\rm H I}$\,$\rangle$ (zero flux)\\
		\hline 
		6.43 &8921-9139& Ly$\alpha$ & 0.94$\pm$0.08 & 0.78$\pm$0.07\\
		6.25 &8711-8921& Ly$\alpha$ & 0.96$\pm$0.08 & 0.81$\pm$0.08\\
		6.08 &8508-8711& Ly$\alpha$ & 0.94$\pm$0.08 & 0.78$\pm$0.08\\
		5.92 &8312-8508& Ly$\alpha$ & 1.00$\pm$0.09 & 0.91$\pm$0.09\\
		5.76 &8123-8312& Ly$\alpha$ & 0.94$\pm$0.09 &  0.56$\pm$0.07\\
		5.61 &7939-8123& Ly$\alpha$ & 0.94$\pm$0.09 & 0.72$\pm$0.08\\
		\hline
		6.45 &7548-7732& Ly$\beta$ & 0.98$\pm$0.09 &0.97$\pm$0.09\\
		\hline
	\end{tabular}
\end{table*}

\section{Discussion}

With the medium-resolution spectrum of the target, we perform three different absorption tests to understand cosmic reionization. We extend the measurement of the Gunn-Peterson transmission, dark gap, and dark pixel statistics to $\mathnormal{z}$\,$\sim$6.5, where constraints are still poor. The results are consistent with the sudden change around 5.8$\leq$\,$\mathnormal{z}$\,$\leq$6.1, suggesting the end stage of the reionization epoch at the redshift interval, which is also consistent with previous studies \citep{Fan2006b,Eilers2018,Bosman2018}. Our measurements at  $\mathnormal{z}>$6.18 also indicates a continuing trend of increasing neutral fractions toward higher redshifts. However, we find a subtle difference in our results. 
\\
For IGM transmission measurement, the factors affecting the measurement come from the different ways to extrapolate the intrinsic continuum to the blueward of Ly$\alpha$ emission, the LOS variance, and the data quality. 
First, there are two main ways to extrapolate the continuum to the blue of the Ly$\alpha$ emission line, using the power-law continuum (or power-law plus Ly$\alpha$ and  $N_{\rm V}$\,$\lambda$1239,1243 emissions) \citep[]{Fan2006b, Becker2015, Barnett2017, Bosman2018} or using the principal component analysis (PCA)\citep[]{Mazzucchelli2017, Eilers2018}.  PCA tends to predict higher continuum levels \citep{Eilers2018}, thus the transmissions measured by PCA would be lower. 

~\citet{Eilers2018} and ~\citet{Fan2006b} measured the same sightlines, but found the median transmission difference of -0.023. Because these measurements are in the same sightline, the sources of the difference could be different observations, different data reduction, and different continuum estimation. While ~\citet{Eilers2018} used PCA to estimate the continua, ~\citet{Fan2006b} assumed a power-law cotinua.
Besides, the mean transmission in ~\citet{Eilers2018} is by $\sim$0.02 lower than that measured by ~\citet{Bosman2018} using power-law continuum at 5$<$z$<$6 but through different LOSs. 
However, because these measurements are through different sightlines, it is unclear how much of the difference is due to the systematics, and how much is from the inhomogeneity of the reionization.
In comparing with our measurements, at z$>$6, ~\citet{Eilers2018} measured T$\approx$0 (Fig.~\ref{trans}) using PCA continuum, while the transmission in this work is $\approx$0.011$\pm$0.009, higher but still consistent within 2-$\sigma$, in a different sightline. Because these QSOs are at different sightlines, the difference of 0.011 could be real due to the inhomogeneity in reionization.
In addition, the data quality and data analysis are crucial for the transmission measurement at high-z as well. At z$>$6, the GP absorption can be easily saturated due to the highly opaque IGM. Because of this, only a few faint transmission spikes can be observed. However, this wavelength window is overlapped with sky emissions. Compared with \citet{Tang2017}, because we have a higher S/N and longer exposure time spectrum in this work, the sky contamination is weaker and the transmitted spikes are seen clearer (Fig.~\ref{fractrans}). Instead of masking the sky, \citet{Tang2017} measure the squared-spectrum-error-weighted transmission to reduce the effect of the sky. However, their result is still suffering from the over-subtracted sky. Nevertheless, our result is consistent within 2-$\sigma$. 
The near zone region, measured from the transmission, aside from the diverse property between individual QSOs, also has LOS difference and difference between different data quality and analysis. The  $R_\text{NZ,corrected}$ of this QSO is within the scatter of other  z$\sim$6.5 QSOs. When comparing $R_\text{NZ,corrected}$ of the same QSO measured in different studies, the $R_\text{NZ,corrected}$ we measured is 5.79$\pm$0.09 $p$Mpc, which agrees with the previous measurement of 4.3$\pm$1.5 $p$Mpc by ~\citet{Tang2017} within the uncertainty. However, it is in disagreement with 6.78$\pm$0.09 $p$Mpc by ~\citet{Mazzucchelli2017}. 
We suspect the disagreement is due to the different observations and data analysis; the exposure times were 30 and 83.3 minutes in ~\citet{Mazzucchelli2017} and in ~\citet{Tang2017}, while our spectrum was measured with 7.5 hours of exposure time. The scarcely remaining flux at such wavelengths could lead to the larger uncertainty on the measurements. Moreover, ~\citet{Mazzucchelli2017} used the correction function derived by ~\citet{Eilers2017} for higher redshift QSOs, which is $R_\text{NZ,corrected}$= $R_\text{p,NZ}$\,$\times$\,$10^{0.4(27+M_{1450})/2.35}$. The scale factor is 1.094 times larger than ours, resulting in a closer value of 6.33 $p$Mpc if we multiply our result by 1.094. In addition, the continuum estimation methods were largely different. ~\citet{Mazzucchelli2017} performed the continuum fitting using PCA, the transmission profile at the near zone region depends on how well PCA can reconstruct the intrinsic spectrum. To overcome the shortage of the transmission measurement, we take advantage of the high resolution and high S/N rate spectrum to try two additional methods.

In the dark gap distribution, the gap width development of both the $\tau$\,$>$2.5 gaps and $\tau$\,$>$3.5 gaps in our spectrum follow a similar rising trend as in \citet{Fan2006b} at z$\lesssim$6. However, at z$>$6, the sizes of the $\tau$\,$>$3.5 gaps decrease predominantly due to the presence of the skyline residuals and the near zone region of the QSO. To be consistent with previous studies \citep{Songaila2002, Fan2006b}, we do not attempt to eliminate the skyline residuals. We caution that the dark gap could be sub-split into smaller gaps by the skyline residuals, especially for the $\tau$\,$ >$3.5 gaps.
In Fig.~\ref{fig:cpsgpa}, we measure the $\tau$\,$>$2.5 dark gap distribution at 5.5$<$\,$\mathnormal{z}$\,$<$6.0 and 5.7$<$\,$\mathnormal{z}$\,$<$6.3 in order to compare it with previous observations \citep{Songaila2002} and simulations \citep{Paschos2005, Gallerani2006}. At 5.5$<$\,$\mathnormal{z}$\,$<$6.0, we compare our result to the simulation from \citet{Paschos2005} at the same redshift range but with a finer resolution (R=5300), and the observation from \citet{Songaila2002} at the same smoothed resolution (R=2600) but at lower redshift (5.5$<$\,$\mathnormal{z}$\,$<$6.0). While the gap widths in \citet{Songaila2002} are mostly shorter than 18 \AA, the widths measured in this work extends to $\sim$56 \AA. The redshift evolution of the gap width is clearly seen. When compared with \citet{Paschos2005}, our result is inconsistent with their simulation. The limited boxsize (6.8cMpc) in their simulation was pointed out by \citet{Gallerani2006}. The resolution difference causes the discrepancy as well. As shown in \citet{Paschos2005}, the fraction of the narrower gaps is higher in simulation when the resolution is higher. At 5.7$<$\,$\mathnormal{z}$\,$<$6.3, the resolution of the simulation performed by ~\citet{Gallerani2006} had been smoothed to the resolution similar to the spectral resolution in ~\citet{Songaila2002}. Although our result is consistent within 1-$\sigma$ with simulation, the distribution measured through only one single LOS is insufficient. Observation of more dark gaps at z$>$5.5 is required to draw the final conclusion.
 \\
 For the dark pixel fraction, we extend the measurement to $\mathnormal{z}$\,$\sim$6.5 with this single target. Simply comparing our result to the statistical result shown in \citet{McGreer2011} and \citet{McGreer2015}, the measurement using the 2-$\sigma$ threshold is much higher than that from using the negative pixel threshold. The 2-$\sigma$ threshold method is sensitive to how accurate the per pixel noise can be measured.  In Fig.~\ref{fig:dps} bottom panel (zero flux threshold) we measure the upper limit of$\langle$\,$x_{\rm H I}$\,$\rangle$\,$\sim$ 0.8 at $\mathnormal{z}$\,$\sim$5.8, which is larger than the measurement of \citet{McGreer2015}. This again suggests the requirement for more high redshift QSO spectra, since there is a wide LOS variance at these redshifts.
\\

\section{Summary}

We obtained a deep, medium-resolution optical spectrum of the QSO, PSO J006.1240+39.2219, at $\mathnormal{z}$=6.62 with the Subaru 8m telescope. We measure the line luminosity of Ly$\alpha$ as 17.75$\pm $0.10$\times$ 10$^{44}$erg s$^{-1}$, \ion{N}{5}\,$\lambda$1239,1243 as 1.79$\pm$0.06 $\times$10$^{44}$erg s$^{-1}$ and the \ion{O}{1}\,$\lambda$1304+\,\ion{Si}{2}\,$\lambda$1306 as 1.54$\pm$0.16 $\times$10$^{44}$erg s$^{-1}$. We found a slightly smaller $\tau$ than previous studies at z$>$6. However, $\tau$ is still increasing with increasing redshifts.
The redshift of the sudden change in opacity is consistent with literature,  which is around $\mathnormal{z}$\,$\sim$5.8; The normalized, $R_\text{NZ,corrected}$, is 5.62$\pm$0.09 $p$Mpc, being consistent with decreasing sizes at higher-$\mathnormal{z}$.  We also investigate the distribution of the dark gaps in the spectrum, which also shows a significant increase of the gap widths at $\mathnormal{z}$\,$>$6. Furthermore, we extend the dark pixel measurements to $\mathnormal{z}>$6. Using the zero flux threshold, the measured upper limit at $\mathnormal{z}<$5.8 is $\langle$\,$x_{\rm H I}$\,$\rangle$\,$\sim$0.5, while the upper limit is $\langle$\,$x_{\rm H I}$\,$\rangle$\,$\sim$0.8 at $\mathnormal{z}>$5.8.

\section*{Acknowledgements}
 We thank the anonymous referee for many helpful suggestions. TG acknowledges the support by the Ministry of Science and Technology of Taiwan (MOST) through grant 105-2112-M-007-003-MY3 and 108-2628-M-007-004-MY3. TYL acknowledges the support by the MOST through grant 107-2813-C-007-104-M. AYLO and TH are supported by the Center for Informatics and Computation in Astronomy (CICA) at National Tsing Hua University (NTHU) through a grant from the Ministry of Education of the Republic of China (Taiwan). AYLO’s visit to NTHU is hosted by Prof Albert Kong and supported by the MOST through grant 105-2119-M-007-028-MY3.

%




\end{document}